\documentclass[a4paper]{llncs}
\usepackage{holtexbasic,url}
\newenvironment{holthmenv}{\begin{array}[t]{l}}{\end{array}}
\renewcommand{\HOLTokenTurnstile}{\ensuremath{\vdash\!\!}}
\renewcommand{\HOLKeyword}[1]{\mathsf{#1}}
\renewcommand{\HOLConst}[1]{{\textsf{\upshape #1}}}
\renewcommand{\HOLSymConst}[1]{\HOLConst{#1}}
\renewcommand{\HOLinline}[1]{\ensuremath{#1}}
\newcommand{\ie}{i.e.\ }

\title{Clocked Definitions in HOL}

\author{Ramana Kumar\inst{1} \and Magnus O. Myreen\inst{2}}

\institute{Data61, CSIRO / NICTA / UNSW
\and CSE Department, Chalmers University of Technology, Sweden}

\begin{document}

\maketitle

\begin{abstract}

Many potentially non-terminating functions cannot be directly defined in a logic of total functions, such as HOL.
A well-known solution to this is to define non-terminating functions using a clock that forces termination at a certain depth of evaluation.
Such clocked definitions are often frowned upon and avoided, since the clock is perceived as extra clutter.
In this short paper, we explain that there are different ways to add a clock, some less intrusive than others.
Our contribution is a technique by which termination proofs are kept simple even when minimising the use of the clock mechanism.
Our examples are definitions of semantic interpreters for programming languages, so called functional big-step semantics.

\end{abstract}

\section{Introduction}\label{sec:intro}
Some functions are naturally non-terminating, which makes them non-trivial to model in a logic of total functions.
A prototypical example is a \emph{definitional interpreter}~\cite{DBLP:journals/lisp/Reynolds98a}, which defines the semantics of a programming language and diverges when the program being interpreted should diverge.
A nice way to make a total model of such a function is to add a clock: an extra parameter whose ticks act as fuel for recursive calls.
The clock makes the function terminating, since any application will eventually run out of fuel, without sacrificing reasoning about divergence, since divergence is equivalent to timing out for every initial clock.
The clock idea is well-known to users of ACL2~\cite{DBLP:conf/tphol/KaufmannM08}, and we have recently advocated for its use in higher-order logic (HOL) for definitions of programming-language semantics~\cite{ESOP16}.

There is trade off to be made whenever a clocked function is defined: how intricate should the clock mechanism be?
A more nuanced clock might lead to a better semantics but require a more subtle termination proof.
There are two dimensions to consider:
\begin{itemize}
  \item[1)]On which recursive calls is the clock decremented?
  \item[2)]Does the clock measure the depth or the length of execution? Equivalently: is the clock environment-like (not returned) or state-like (threaded through)?
\end{itemize}

At the simple end of the first dimension is a clock that consumes fuel on every recursive call.
While conceptually neat, and good for proving termination, this kind of clock mechanism can easily become a burden in later proofs.
By contrast, if the clock is decremented only on problematic recursive calls, there is less clock-related overhead in proofs.

Along the second dimension, an environment-like clock supports a straightforward termination proof, whereas a state-like clock can make termination rather tricky.
The difference is akin to reasoning about accumulator-passing versus directly recursive functions.
However, unlike accumulator-passing style, measuring length versus depth of execution with the clock is a real semantic difference, and certain applications may call for one or the other.


In this paper, we illustrate the four options entailed by the two dimensions above and explain how clocked functions in each style can be defined.
Our technical contribution is a simple technique by which the tricky termination proofs arising from a state-like clock are made simple.

One can read this paper as a tutorial on how to neatly define the clocked functions used in our previous paper on functional big-step semantics~\cite{ESOP16}.
The technique in Section~\ref{sec:fixclock} has been used to clean up the definitions of the semantics for the CakeML compiler's 13 intermediate languages (\url{https://cakeml.org}).


\paragraph{\bf Running example.} 

We will use Nipkow and Klein's IMP language from \emph{Concrete Semantics}~\cite{DBLP:books/sp/NipkowK14} as a running example.
The IMP language is a simple \HOLinline{\HOLConst{While}}-language with the following abstract syntax.
\[
  \begin{holthmenv}
    \HOLTyOp{aexp}\;=\;\HOLConst{N}\;\HOLTyOp{int}\;\HOLTokenBar{}\;\HOLConst{V}\;\HOLTyOp{str}\;\HOLTokenBar{}\;\HOLConst{Plus}\;\HOLTyOp{aexp}\;\HOLTyOp{aexp}\\

    \HOLTyOp{bexp}\;=\;\HOLConst{Bc}\;\HOLTyOp{bool}\;\HOLTokenBar{}\;\HOLConst{Not}\;\HOLTyOp{bexp}\;\HOLTokenBar{}\;\HOLConst{And}\;\HOLTyOp{bexp}\;\HOLTyOp{bexp}\;\HOLTokenBar{}\;\HOLConst{Less}\;\HOLTyOp{aexp}\;\HOLTyOp{aexp}\\

    \HOLTyOp{com}\;=\;\HOLConst{Skip}\;\HOLTokenBar{}\;\HOLConst{Set}\;\HOLTyOp{str}\;\HOLTyOp{aexp}\;\HOLTokenBar{}\;\HOLConst{Seq}\;\HOLTyOp{com}\;\HOLTyOp{com}\;\HOLTokenBar{}\;\HOLConst{If}\;\HOLTyOp{bexp}\;\HOLTyOp{com}\;\HOLTyOp{com}\;\HOLTokenBar{}\;\HOLConst{While}\;\HOLTyOp{bexp}\;\HOLTyOp{com}
  \end{holthmenv}
\]
Arithmetic and Boolean expressions are given semantics (\HOLinline{\HOLConst{aval}} and \HOLinline{\HOLConst{bval}} respectively) as would be expected, as functions.

In what follows we define four clocked functional semantics for this language. Section~\ref{sec:environment} gives definitions with environment-like clocks, and shows the pros and cons, within this style, of using the clock on every recursive call.
Section~\ref{sec:state} describes, and contrasts, state-like clocks, and illustrates the tricky termination problems they produce and how to solve them.

\section{Environment-like clocks}\label{sec:environment}\label{sec:ev}

Our first evaluation function for commands has a simple clock mechanism: we decrement the clock on every recursive call, and do not return it.
The clock, \HOLinline{\HOLFreeVar{t}}, is a natural number, and the result of evaluation is either \HOLinline{\HOLConst{None}} representing timeout or \HOLinline{\HOLConst{Some}\;\HOLFreeVar{s}} with the final state.
\[
  \begin{holthmenv}
    \HOLConst{ev}\;\HOLFreeVar{c}\;\HOLFreeVar{s}\;\HOLNumLit{0}\;\HOLSymConst{=}\;\HOLConst{None}\\
\HOLConst{ev}\;\HOLConst{Skip}\;\HOLFreeVar{s}\;(\HOLConst{Suc}\;\HOLFreeVar{t})\;\HOLSymConst{=}\;\HOLConst{Some}\;\HOLFreeVar{s}\\
\HOLConst{ev}\;(\HOLConst{Set}\;\HOLFreeVar{x}\;\HOLFreeVar{a})\;\HOLFreeVar{s}\;(\HOLConst{Suc}\;\HOLFreeVar{t})\;\HOLSymConst{=}\;\HOLConst{Some}\;((\HOLFreeVar{x}\;\HOLSymConst{\ensuremath{\mapsto}}\;\HOLConst{aval}\;\HOLFreeVar{a}\;\HOLFreeVar{s})\;\HOLFreeVar{s})\\
\HOLConst{ev}\;(\HOLConst{Seq}\;\HOLFreeVar{c\sb{\mathrm{1}}}\;\HOLFreeVar{c\sb{\mathrm{2}}})\;\HOLFreeVar{s}\;(\HOLConst{Suc}\;\HOLFreeVar{t})\;\HOLSymConst{=}\\
\;\;\HOLKeyword{case}\;\HOLConst{ev}\;\HOLFreeVar{c\sb{\mathrm{1}}}\;\HOLFreeVar{s}\;\HOLFreeVar{t}\;\HOLKeyword{of}\;\HOLConst{None}\;\HOLTokenImp{}\;\HOLConst{None}\;\HOLTokenBar{}\;\HOLConst{Some}\;\HOLBoundVar{s\sb{\mathrm{2}}}\;\HOLTokenImp{}\;\HOLConst{ev}\;\HOLFreeVar{c\sb{\mathrm{2}}}\;\HOLBoundVar{s\sb{\mathrm{2}}}\;\HOLFreeVar{t}\\
\HOLConst{ev}\;(\HOLConst{If}\;\HOLFreeVar{b}\;\HOLFreeVar{ct}\;\HOLFreeVar{cf})\;\HOLFreeVar{s}\;(\HOLConst{Suc}\;\HOLFreeVar{t})\;\HOLSymConst{=}\;\HOLConst{ev}\;(\HOLKeyword{if}\;\HOLConst{bval}\;\HOLFreeVar{b}\;\HOLFreeVar{s}\;\HOLKeyword{then}\;\HOLFreeVar{ct}\;\HOLKeyword{else}\;\HOLFreeVar{cf})\;\HOLFreeVar{s}\;\HOLFreeVar{t}\\
\HOLConst{ev}\;(\HOLConst{While}\;\HOLFreeVar{g}\;\HOLFreeVar{c})\;\HOLFreeVar{s}\;(\HOLConst{Suc}\;\HOLFreeVar{t})\;\HOLSymConst{=}\\
\;\;\HOLKeyword{if}\;\HOLConst{bval}\;\HOLFreeVar{g}\;\HOLFreeVar{s}\;\HOLKeyword{then}\;\HOLConst{ev}\;(\HOLConst{Seq}\;\HOLFreeVar{c}\;(\HOLConst{While}\;\HOLFreeVar{g}\;\HOLFreeVar{c}))\;\HOLFreeVar{s}\;\HOLFreeVar{t}\;\HOLKeyword{else}\;\HOLConst{Some}\;\HOLFreeVar{s}
  \end{holthmenv}
\]
This definition may look neat at first, but observe that most clauses require a non-zero input clock (\HOLinline{\HOLConst{Suc}\;\HOLFreeVar{t}}), which means that case analysis on the clock is required in all proofs.
Nevertheless, the termination proof is trivial, since the clock itself is a well-founded measure. (The termination proof is automatic in the HOL theorem prover.)

The function above can easily be modified to not check the clock for non-recursive cases, e.g. \HOLinline{\HOLConst{Skip}}, but the result is just a less well presented function.

\subsection{Problem with decrement-everywhere style}

The problem with functions that decrement the clock on every recursive call is that nearly every proof needs to assume a lower limit on the clock or use induction.
As a simple example, consider proving that \HOLinline{\HOLConst{Seq}\;(\HOLConst{Seq}\;\HOLFreeVar{p}\;\HOLConst{Skip})\;(\HOLConst{Seq}\;\HOLConst{Skip}\;\HOLFreeVar{q})} is the same as \HOLinline{\HOLConst{Seq}\;\HOLFreeVar{p}\;\HOLFreeVar{q}}.
The proof of this simple property is made very cumbersome by the eager use of the clock in \HOLinline{\HOLConst{ev}}.

To prove a theorem relating the evaluation of the two commands,
we require some very specific assumptions about the clock:
\[
  \begin{holthmenv}
    \HOLTokenTurnstile{}\;(\HOLSymConst{\HOLTokenForall{}}\HOLBoundVar{s}.\;\HOLConst{ev}\;\HOLFreeVar{p}\;\HOLBoundVar{s}\;(\HOLFreeVar{t}\;\HOLSymConst{\ensuremath{-}}\;\HOLNumLit{1})\;\HOLSymConst{=}\;\HOLConst{ev}\;\HOLFreeVar{p}\;\HOLBoundVar{s}\;(\HOLFreeVar{t}\;\HOLSymConst{\ensuremath{-}}\;\HOLNumLit{2}))\;\HOLSymConst{\HOLTokenConj{}}\\
\;\;\;(\HOLSymConst{\HOLTokenForall{}}\HOLBoundVar{s}.\;\HOLConst{ev}\;\HOLFreeVar{q}\;\HOLBoundVar{s}\;(\HOLFreeVar{t}\;\HOLSymConst{\ensuremath{-}}\;\HOLNumLit{1})\;\HOLSymConst{=}\;\HOLConst{ev}\;\HOLFreeVar{q}\;\HOLBoundVar{s}\;(\HOLFreeVar{t}\;\HOLSymConst{\ensuremath{-}}\;\HOLNumLit{2}))\;\HOLSymConst{\HOLTokenConj{}}\;\HOLNumLit{2}\;\HOLSymConst{\HOLTokenLt{}}\;\HOLFreeVar{t}\;\HOLSymConst{\HOLTokenImp{}}\\
\;\;\;\HOLConst{ev}\;(\HOLConst{Seq}\;(\HOLConst{Seq}\;\HOLFreeVar{p}\;\HOLConst{Skip})\;(\HOLConst{Seq}\;\HOLConst{Skip}\;\HOLFreeVar{q}))\;\HOLFreeVar{s}\;\HOLFreeVar{t}\;\HOLSymConst{=}\;\HOLConst{ev}\;(\HOLConst{Seq}\;\HOLFreeVar{p}\;\HOLFreeVar{q})\;\HOLFreeVar{s}\;\HOLFreeVar{t}
  \end{holthmenv}
\]
These assumptions are a nuisance to deal with in any proof where we want to use the theorem above.

\subsection{Minimal use of the clock}\label{sec:minimal}

It is preferable to decrement the clock only on problematic recursive calls, rather than on every call as above.
In our example, the only problematic call is the recursion in the \HOLinline{\HOLConst{While}} case.
In all other cases, the expression that is being evaluated shrinks.
Therefore, we can define the clocked function as follows with a clock-check-and-decrement only in the \HOLinline{\HOLConst{While}} case.
\[
  \begin{holthmenv}
    \HOLConst{ev_min}\;\HOLConst{Skip}\;\HOLFreeVar{s}\;\HOLFreeVar{t}\;\HOLSymConst{=}\;\HOLConst{Some}\;\HOLFreeVar{s}\\
\HOLConst{ev_min}\;(\HOLConst{Set}\;\HOLFreeVar{x}\;\HOLFreeVar{a})\;\HOLFreeVar{s}\;\HOLFreeVar{t}\;\HOLSymConst{=}\;\HOLConst{Some}\;((\HOLFreeVar{x}\;\HOLSymConst{\ensuremath{\mapsto}}\;\HOLConst{aval}\;\HOLFreeVar{a}\;\HOLFreeVar{s})\;\HOLFreeVar{s})\\
\HOLConst{ev_min}\;(\HOLConst{Seq}\;\HOLFreeVar{c\sb{\mathrm{1}}}\;\HOLFreeVar{c\sb{\mathrm{2}}})\;\HOLFreeVar{s}\;\HOLFreeVar{t}\;\HOLSymConst{=}\\
\;\;\HOLKeyword{case}\;\HOLConst{ev_min}\;\HOLFreeVar{c\sb{\mathrm{1}}}\;\HOLFreeVar{s}\;\HOLFreeVar{t}\;\HOLKeyword{of}\;\HOLConst{None}\;\HOLTokenImp{}\;\HOLConst{None}\;\HOLTokenBar{}\;\HOLConst{Some}\;\HOLBoundVar{s\sb{\mathrm{2}}}\;\HOLTokenImp{}\;\HOLConst{ev_min}\;\HOLFreeVar{c\sb{\mathrm{2}}}\;\HOLBoundVar{s\sb{\mathrm{2}}}\;\HOLFreeVar{t}\\
\HOLConst{ev_min}\;(\HOLConst{If}\;\HOLFreeVar{b}\;\HOLFreeVar{c\sb{\mathrm{1}}}\;\HOLFreeVar{c\sb{\mathrm{2}}})\;\HOLFreeVar{s}\;\HOLFreeVar{t}\;\HOLSymConst{=}\;\HOLConst{ev_min}\;(\HOLKeyword{if}\;\HOLConst{bval}\;\HOLFreeVar{b}\;\HOLFreeVar{s}\;\HOLKeyword{then}\;\HOLFreeVar{c\sb{\mathrm{1}}}\;\HOLKeyword{else}\;\HOLFreeVar{c\sb{\mathrm{2}}})\;\HOLFreeVar{s}\;\HOLFreeVar{t}\\
\HOLConst{ev_min}\;(\HOLConst{While}\;\HOLFreeVar{b}\;\HOLFreeVar{c})\;\HOLFreeVar{s}\;\HOLFreeVar{t}\;\HOLSymConst{=}\\
\;\;\HOLKeyword{if}\;\HOLConst{bval}\;\HOLFreeVar{b}\;\HOLFreeVar{s}\;\HOLKeyword{then}\\
\;\;\;\;\HOLKeyword{if}\;\HOLFreeVar{t}\;\HOLSymConst{=}\;\HOLNumLit{0}\;\HOLKeyword{then}\;\HOLConst{None}\;\HOLKeyword{else}\;\HOLConst{ev_min}\;(\HOLConst{Seq}\;\HOLFreeVar{c}\;(\HOLConst{While}\;\HOLFreeVar{b}\;\HOLFreeVar{c}))\;\HOLFreeVar{s}\;(\HOLFreeVar{t}\;\HOLSymConst{\ensuremath{-}}\;\HOLNumLit{1})\\
\;\;\HOLKeyword{else}\;\HOLConst{Some}\;\HOLFreeVar{s}
  \end{holthmenv}
\]

With such a definition of the semantics, it is easy to prove that \HOLinline{\HOLConst{Seq}\;\HOLFreeVar{p}\;\HOLFreeVar{q}} always evaluates the same as \HOLinline{\HOLConst{Seq}\;(\HOLConst{Seq}\;\HOLFreeVar{p}\;\HOLConst{Skip})\;(\HOLConst{Seq}\;\HOLConst{Skip}\;\HOLFreeVar{q})}.
The equality can be stated without assumptions and used directly as a rewrite rule.
\[
  \begin{holthmenv}
    \HOLTokenTurnstile{}\;\HOLConst{ev_min}\;(\HOLConst{Seq}\;(\HOLConst{Seq}\;\HOLFreeVar{p}\;\HOLConst{Skip})\;(\HOLConst{Seq}\;\HOLConst{Skip}\;\HOLFreeVar{q}))\;\HOLFreeVar{s}\;\HOLFreeVar{t}\;\HOLSymConst{=}\;\HOLConst{ev_min}\;(\HOLConst{Seq}\;\HOLFreeVar{p}\;\HOLFreeVar{q})\;\HOLFreeVar{s}\;\HOLFreeVar{t}
  \end{holthmenv}
\]

\paragraph{\bf Termination proof.}

The equations defining \HOLinline{\HOLConst{ev_min}} terminate because the lexicographic combination of measuring the clock and the size of the evaluated expression is well-founded.
For each recursive call, the clock either stays the same or shrinks; if it stays the same, then the size of the expression shrinks.

\section{State-like clocks}\label{sec:state}

The previous section defined functions where the clock limits the depth of evaluation.
In some circumstances, e.g. when defining interpreters, it makes sense to limit the \emph{length} of evaluation instead.
This length is related to the length of an equivalent trace in a small-step semantics.
The difference can be seen most clearly in constructs, like \HOLinline{\HOLConst{Seq}\;\HOLFreeVar{c\sb{\mathrm{1}}}\;\HOLFreeVar{c\sb{\mathrm{2}}}}, where order might matter: we can either give clock ticks to each sub-expression independently, or we can use the same ticks for both and, say, only evaluate \HOLinline{\HOLFreeVar{c\sb{\mathrm{2}}}} after \HOLinline{\HOLFreeVar{c\sb{\mathrm{1}}}}'s ticks have been subtracted.

In the following definition, the clock is passed around as state, limiting length rather than depth.
Each \HOLinline{\HOLConst{Some}}-result contains a store-and-clock pair \HOLinline{(\HOLFreeVar{s}\HOLSymConst{,}\HOLFreeVar{t})}.
\[
  \begin{holthmenv}
    \HOLConst{cval}\;\HOLConst{Skip}\;\HOLFreeVar{s}\;\HOLFreeVar{t}\;\HOLSymConst{=}\;\HOLConst{Some}\;(\HOLFreeVar{s}\HOLSymConst{,}\HOLFreeVar{t})\\
\HOLConst{cval}\;(\HOLConst{Set}\;\HOLFreeVar{x}\;\HOLFreeVar{a})\;\HOLFreeVar{s}\;\HOLFreeVar{t}\;\HOLSymConst{=}\;\HOLConst{Some}\;((\HOLFreeVar{x}\;\HOLSymConst{\ensuremath{\mapsto}}\;\HOLConst{aval}\;\HOLFreeVar{a}\;\HOLFreeVar{s})\;\HOLFreeVar{s}\HOLSymConst{,}\HOLFreeVar{t})\\
\HOLConst{cval}\;(\HOLConst{Seq}\;\HOLFreeVar{c\sb{\mathrm{1}}}\;\HOLFreeVar{c\sb{\mathrm{2}}})\;\HOLFreeVar{s}\;\HOLFreeVar{t}\;\HOLSymConst{=}\\
\;\;\HOLKeyword{case}\;\HOLConst{cval}\;\HOLFreeVar{c\sb{\mathrm{1}}}\;\HOLFreeVar{s}\;\HOLFreeVar{t}\;\HOLKeyword{of}\;\HOLConst{None}\;\HOLTokenImp{}\;\HOLConst{None}\;\HOLTokenBar{}\;\HOLConst{Some}\;(\HOLBoundVar{s\sb{\mathrm{2}}}\HOLSymConst{,}\HOLBoundVar{t\sb{\mathrm{2}}})\;\HOLTokenImp{}\;\HOLConst{cval}\;\HOLFreeVar{c\sb{\mathrm{2}}}\;\HOLBoundVar{s\sb{\mathrm{2}}}\;\HOLBoundVar{t\sb{\mathrm{2}}}\\
\HOLConst{cval}\;(\HOLConst{If}\;\HOLFreeVar{b}\;\HOLFreeVar{c\sb{\mathrm{1}}}\;\HOLFreeVar{c\sb{\mathrm{2}}})\;\HOLFreeVar{s}\;\HOLFreeVar{t}\;\HOLSymConst{=}\;\HOLConst{cval}\;(\HOLKeyword{if}\;\HOLConst{bval}\;\HOLFreeVar{b}\;\HOLFreeVar{s}\;\HOLKeyword{then}\;\HOLFreeVar{c\sb{\mathrm{1}}}\;\HOLKeyword{else}\;\HOLFreeVar{c\sb{\mathrm{2}}})\;\HOLFreeVar{s}\;\HOLFreeVar{t}\\
\HOLConst{cval}\;(\HOLConst{While}\;\HOLFreeVar{b}\;\HOLFreeVar{c})\;\HOLFreeVar{s}\;\HOLFreeVar{t}\;\HOLSymConst{=}\\
\;\;\HOLKeyword{if}\;\HOLConst{bval}\;\HOLFreeVar{b}\;\HOLFreeVar{s}\;\HOLKeyword{then}\\
\;\;\;\;\HOLKeyword{if}\;\HOLFreeVar{t}\;\HOLSymConst{=}\;\HOLNumLit{0}\;\HOLKeyword{then}\;\HOLConst{None}\;\HOLKeyword{else}\;\HOLConst{cval}\;(\HOLConst{Seq}\;\HOLFreeVar{c}\;(\HOLConst{While}\;\HOLFreeVar{b}\;\HOLFreeVar{c}))\;\HOLFreeVar{s}\;(\HOLFreeVar{t}\;\HOLSymConst{\ensuremath{-}}\;\HOLNumLit{1})\\
\;\;\HOLKeyword{else}\;\HOLConst{Some}\;(\HOLFreeVar{s}\HOLSymConst{,}\HOLFreeVar{t})
  \end{holthmenv}
\]

\subsection{Challenging termination proof}

Proving termination for functions that treat the clock as state is not as straightforward as previously.
The reason for termination is the same as for \HOLinline{\HOLConst{ev_min}} above, i.e. the clock decreases when the size of the expression does not.
However, the termination proof is more difficult because the clock comes from recursive calls rather than from input arguments.

In the IMP language, this problem shows up in the termination goal for \HOLinline{\HOLConst{cval}\;(\HOLConst{Seq}\;\HOLFreeVar{c\sb{\mathrm{1}}}\;\HOLFreeVar{c\sb{\mathrm{2}}})\;\HOLFreeVar{s}\;\HOLFreeVar{t}}: here one needs to show that evaluation of \HOLinline{\HOLFreeVar{c\sb{\mathrm{2}}}} in the state and clock produced by \HOLinline{\HOLConst{cval}\;\HOLFreeVar{c\sb{\mathrm{1}}}\;\HOLFreeVar{s}\;\HOLFreeVar{t}} is smaller than the original input, \ie \HOLinline{\HOLConst{Seq}\;\HOLFreeVar{c\sb{\mathrm{1}}}\;\HOLFreeVar{c\sb{\mathrm{2}}}} and \HOLinline{\HOLFreeVar{t}}.
This goal is problematic because we have yet to define \HOLinline{\HOLConst{cval}}. Though technically possible with modern definition packages, it is cumbersome to prove lemmas about \HOLinline{\HOLConst{cval}} before its termination proof is complete.

A common trick to avoid such difficult termination proofs is to define a function with redundant safety checks that make the termination proof simple.
Once the function is defined, we can manually prove definition-like equations without the added safety checks.
The safety checks also need to be removed from the induction theorems produced for these definitions.

The obvious way to instrument a definition with checks that make the termination proof simple is to add a redundant safety check on the arguments to any problematic recursive call.
For example, we could rephrase the problematic \HOLinline{\HOLConst{Seq}} case as follows with a redundant check of \HOLinline{\HOLFreeVar{t}\;\HOLSymConst{\HOLTokenLt{}}\;\HOLFreeVar{t\sb{\mathrm{2}}}}.
We know that this check should always be false, but it is difficult to establish that property in the middle of the termination proof.
\[
  \begin{holthmenv}
    \HOLConst{cval}\;(\HOLConst{Seq}\;\HOLFreeVar{c\sb{\mathrm{1}}}\;\HOLFreeVar{c\sb{\mathrm{2}}})\;\HOLFreeVar{s}\;\HOLFreeVar{t}\;\HOLSymConst{=}\\
\;\;\HOLKeyword{case}\;\HOLConst{cval}\;\HOLFreeVar{c\sb{\mathrm{1}}}\;\HOLFreeVar{s}\;\HOLFreeVar{t}\;\HOLKeyword{of}\\
\;\;\;\;\HOLConst{None}\;\HOLTokenImp{}\;\HOLConst{None}\\
\;\;\HOLTokenBar{}\;\HOLConst{Some}\;(\HOLBoundVar{s\sb{\mathrm{2}}}\HOLSymConst{,}\HOLBoundVar{t\sb{\mathrm{2}}})\;\HOLTokenImp{}\;\HOLConst{cval}\;\HOLFreeVar{c\sb{\mathrm{2}}}\;\HOLBoundVar{s\sb{\mathrm{2}}}\;(\HOLKeyword{if}\;\HOLFreeVar{t}\;\HOLSymConst{\HOLTokenLt{}}\;\HOLBoundVar{t\sb{\mathrm{2}}}\;\HOLKeyword{then}\;\HOLFreeVar{t}\;\HOLKeyword{else}\;\HOLBoundVar{t\sb{\mathrm{2}}})
  \end{holthmenv}
\]
Once the function, in this case \HOLinline{\HOLConst{cval}}, is defined, one can manually prove the desired defining equation for the \HOLinline{\HOLConst{Seq}} case of \HOLinline{\HOLConst{cval}}:
\[
  \begin{holthmenv}
    \HOLConst{cval}\;(\HOLConst{Seq}\;\HOLFreeVar{c\sb{\mathrm{1}}}\;\HOLFreeVar{c\sb{\mathrm{2}}})\;\HOLFreeVar{s}\;\HOLFreeVar{t}\;\HOLSymConst{=}\\
\;\;\HOLKeyword{case}\;\HOLConst{cval}\;\HOLFreeVar{c\sb{\mathrm{1}}}\;\HOLFreeVar{s}\;\HOLFreeVar{t}\;\HOLKeyword{of}\;\HOLConst{None}\;\HOLTokenImp{}\;\HOLConst{None}\;\HOLTokenBar{}\;\HOLConst{Some}\;(\HOLBoundVar{s\sb{\mathrm{2}}}\HOLSymConst{,}\HOLBoundVar{t\sb{\mathrm{2}}})\;\HOLTokenImp{}\;\HOLConst{cval}\;\HOLFreeVar{c\sb{\mathrm{2}}}\;\HOLBoundVar{s\sb{\mathrm{2}}}\;\HOLBoundVar{t\sb{\mathrm{2}}}
  \end{holthmenv}
\]

Although, this technique of adding redundant safety checks to the incoming clock argument works, our experience with the CakeML compiler suggests that the removal of redundant safety checks on inputs tends to be ad hoc and tedious.

\subsection{Simple definition technique: fix-clock wrapper}\label{sec:fixclock}

In the course of defining many clocked functions for CakeML, we realised how the safety checks can be expressed in a way that makes them very easy to remove.

The trick is to perform the safety checks on the \emph{return} from recursive calls (\ie the production of potentially bad values) instead of at the sites of consumption of potentially bad values.
For the running example, we define a new function, \HOLinline{\HOLConst{fix_clock}}, and wrap it around the producer of potentially bad values.
The \HOLinline{\HOLConst{fix_clock}} wrapper adjusts the clock back to its original value if the clock somehow increased during the recursive call:
\[
  \begin{holthmenv}
    \HOLConst{fix_clock}\;\HOLFreeVar{t}\;\HOLConst{None}\;\HOLSymConst{=}\;\HOLConst{None}\\
\HOLConst{fix_clock}\;\HOLFreeVar{t}\;(\HOLConst{Some}\;(\HOLFreeVar{s}\HOLSymConst{,}\HOLFreeVar{t\sp{\prime}}))\;\HOLSymConst{=}\;\HOLKeyword{if}\;\HOLFreeVar{t}\;\HOLSymConst{\HOLTokenLt{}}\;\HOLFreeVar{t\sp{\prime}}\;\HOLKeyword{then}\;\HOLConst{Some}\;(\HOLFreeVar{s}\HOLSymConst{,}\HOLFreeVar{t})\;\HOLKeyword{else}\;\HOLConst{Some}\;(\HOLFreeVar{s}\HOLSymConst{,}\HOLFreeVar{t\sp{\prime}})\\[0.7em]
    \HOLConst{cval}\;(\HOLConst{Seq}\;\HOLFreeVar{c\sb{\mathrm{1}}}\;\HOLFreeVar{c\sb{\mathrm{2}}})\;\HOLFreeVar{s}\;\HOLFreeVar{t}\;\HOLSymConst{=}\\
\;\;\HOLKeyword{case}\;\HOLConst{fix_clock}\;\HOLFreeVar{t}\;(\HOLConst{cval}\;\HOLFreeVar{c\sb{\mathrm{1}}}\;\HOLFreeVar{s}\;\HOLFreeVar{t})\;\HOLKeyword{of}\\
\;\;\;\;\HOLConst{None}\;\HOLTokenImp{}\;\HOLConst{None}\\
\;\;\HOLTokenBar{}\;\HOLConst{Some}\;(\HOLBoundVar{s\sb{\mathrm{2}}}\HOLSymConst{,}\HOLBoundVar{t\sb{\mathrm{2}}})\;\HOLTokenImp{}\;\HOLConst{cval}\;\HOLFreeVar{c\sb{\mathrm{2}}}\;\HOLBoundVar{s\sb{\mathrm{2}}}\;\HOLBoundVar{t\sb{\mathrm{2}}}
  \end{holthmenv}
\]

The formulation above makes the safety checks (\HOLinline{\HOLConst{fix_clock}}) very easy to remove.
Once the \HOLinline{\HOLConst{cval}} function is defined, we prove that the clock never increases
\[
  \begin{holthmenv}
    \HOLTokenTurnstile{}\;\HOLConst{cval}\;\HOLFreeVar{c}\;\HOLFreeVar{s}\;\HOLFreeVar{t}\;\HOLSymConst{=}\;\HOLConst{Some}\;(\HOLFreeVar{s\sp{\prime}}\HOLSymConst{,}\HOLFreeVar{t\sp{\prime}})\;\HOLSymConst{\HOLTokenImp{}}\;\HOLFreeVar{t\sp{\prime}}\;\HOLSymConst{\HOLTokenLeq{}}\;\HOLFreeVar{t}
  \end{holthmenv}
\]
and use this property to prove the following rewrite rule, which we apply to both the defining theorem for \HOLinline{\HOLConst{cval}} and its associated induction theorem.
\[
  \begin{holthmenv}
    \HOLTokenTurnstile{}\;\HOLConst{fix_clock}\;\HOLFreeVar{t}\;(\HOLConst{cval}\;\HOLFreeVar{c}\;\HOLFreeVar{s}\;\HOLFreeVar{t})\;\HOLSymConst{=}\;\HOLConst{cval}\;\HOLFreeVar{c}\;\HOLFreeVar{s}\;\HOLFreeVar{t}
  \end{holthmenv}
\]

\subsection{Decrement everywhere and state-like clock}

The final combination of the two options from the introduction is to pick decrement-everywhere and state-like clock.
We omit such a definition since it is an obvious rephrasing of the decrement-everywhere function \HOLinline{\HOLConst{ev}} from Section~\ref{sec:ev}.

Is this combination useful?
We envision that such formulations can be convenient stepping stones when proving equivalence between functional big-step semantics and small-step semantics, since a length-limiting clock decremented on each call can be made to match the length of the small-step trace.
However, we did not use this combination when proving such an equivalence~\cite{ESOP16}.

\section{Summary and related work}\label{sec:conclusion}

This short paper presents different styles of clocked definitions, and provides a simple definition technique for functions that treat the clock as a state component which gets threaded through evaluation.

Clocked functions are used in all major theorem provers, but are often seen as unwanted and the clocks are considered a burden.
This paper's purpose is to continue our recent work on programming language semantics~\cite{ESOP16} and to show that it is easy to define HOL functions with the clock as a state component.

ACL2 is a prover where clocked functions are encouraged and well supported.
The ACL2 code base is full of examples of clocked functions, e.g. Centaur Inc's industrial Verilog preprocessor\footnote{\url{https://github.com/acl2/acl2/tree/master/books/centaur/vl/loader/preprocessor}, retrieved 2016-03-07}.
ACL2 even supports tricks that allow execution of clocked functions as if they didn't have a clock\footnote{\url{http://www.cs.utexas.edu/users/moore/acl2/vstte-2012/acl2-dkms/problem5/breadth-first.lisp}, retrieved 2016-03-07}.

\paragraph{Acknowledgements.}

We thank Jared Davis for a long list of pointers on how clocked functions are used in ACL2.
NICTA is funded by the Australian Government through the Department
of Communications and the Australian Research Council through the
ICT Centre of Excellence Program.
\bibliographystyle{splncs03}
\bibliography{paper}

\end{document}